\documentclass[11pt]{article}
\usepackage{graphicx}
\usepackage{epsfig}
\usepackage{rotating}

\newcommand{\BABARPubYear}    {06}

\newcommand{\BABARConfNumber} {015}
\newcommand{\SLACPubNumber} {12005}

\newcommand{\pmff}[2]{{}^{+#1}_{-#2}{}_{\mathrm{FF}}}
\newcommand{\DBR}{\mbox {$\Delta \ensuremath{\cal B}$}}
\def\BR {\ensuremath{\cal B}\xspace}

\input babarsym

\def\ulnu{$b \rightarrow u \ell \nu$}
\def\clnu{$b \rightarrow c \ell \nu$}
\def\xulnu{$B \rightarrow X_u \ell \nu$}

\def\pimunu{$B^{0} \rightarrow \pi^{-} \mu^{+} \nu$}
\def\pilnu{$B^{0} \rightarrow \pi^{-} \ell^{+} \nu$}

\def\rholnu{$B \rightarrow \rho \ell \nu$}
\def\dstrlnu{$B \rightarrow D^* \ell \nu$}
\def\bfpilnu{${\BR}(B^{0} \rightarrow \pi^{-} \ell^{+} \nu)$}
\def\bfpilnuq{$\Delta {\BR}(B^{0} \rightarrow \pi^{-} \ell^{+} \nu,q^2)$}
\def\bfpilnuqp{$\DBR(q^2)/\BR$}
\def\bfpilnuqq{$\DBR(q^2)$}

\def\bfpilnuqi{$\DBR(\ensuremath{q^2_i})$}

\def\bfpilnuqj{$\DBR(\ensuremath{q^2_j})$}
\def\lnrt{loose neutrino reconstruction technique}
\def\lnr{loose neutrino reconstruction}
\def\qq{$q^2$}
\def\qqr{$\tilde{q}^2$} 
\def\fplus{$f^+(q^2)$}
\def\vub{$|V_{ub}|$}
\def\nQ{12}
\def\bbz{$B^0\bar{B^0}$}
\def\bpbm{$B^+B^-$}
\def\udsc{$u\bar{u}/d\bar{d}/s\bar{s}/c\bar{c}$}
\def\tautau{$\tau^+\tau^-$}

\def\obb{other $B\bar{B}$}
\def\sigY{$5047 \pm 251$}
\def\ulnY{$10015 \pm 548$}
\def\obbY{$32788 \pm 445$}
\def\conY{$9801 \pm 467$}

\def\cosBYDef{\ensuremath{\cos \theta_{BY} = \left(2 E^*_B
      E^*_Y-m_B^2-m_Y^2\right)/\left(2|\vec p^{\,*}_B|
      |\vec p^{\,*}_Y|\right)}\xspace}

\def\DeltaEDef{\ensuremath{\Delta E = (p_B \cdot p_{\rm beams} - s/2) / 
\sqrt{s}}\xspace}
\def\FitRegion{\ensuremath{|\Delta E| < 1.0~\gev \mbox{ and } m_{\rm ES} > 
5.19~\gevcc}\xspace}

\newcommand{\gevccsq}{\ensuremath{{\mathrm{\,Ge\kern -0.1em V^2\!/}c^4}}\xspace}

\long\def\inst#1{\par\nobreak\kern 4pt\nobreak
    {\it #1}\par\vskip 10pt plus 3pt minus 3pt}

\begin{document}
{\pagestyle{empty}

\begin{flushright}
\babar-CONF-\BABARPubYear/\BABARConfNumber \\
SLAC-PUB-\SLACPubNumber \\
%hep-ex/\LANLNumber \\
\today\ \\
\end{flushright}

\par\vskip 5cm

\begin{center}
\Large \bf Measurement of the \pilnu\ Form Factor Shape and Branching Fraction,
and Determination of $|V_{ub}|$ with a Loose Neutrino Reconstruction Technique
\end{center}
\bigskip

\begin{center}
\large The \babar\ Collaboration\\
\mbox{ }\\
\today
\end{center}
\bigskip \bigskip

\begin{center}
\large \bf Abstract
\end{center}
 We report the results of a study of the exclusive charmless semileptonic \pilnu\
decay undertaken with approximately 227 million $B\bar{B}$ pairs collected at 
the \FourS\ resonance with the \babar\ detector. The analysis uses events in 
which the signal $B$ mesons are reconstructed with a novel \lnrt. We obtain 
partial branching fractions in \nQ\ bins of \qq, the $\ell^+ \nu$ invariant 
mass squared, from which we extract the \fplus\ form factor shape and the total 
branching fraction: \bfpilnu\ $= (1.44 \pm 0.08_{stat} \pm 0.10_{syst})\times 
10^{-4}$. Based on a recent theoretical calculation of the form factor, we find 
the magnitude of the CKM matrix element \vub\ to be $\left(4.1 \pm 0.2_{stat} \pm
0.2_{syst}\pmff{0.6}{0.4} \right) \times {10^{-3}}$, where the last uncertainty 
is due to the normalization of the form factor.  

\vfill
\begin{center}

Submitted to the 33$^{\rm rd}$ International Conference on High-Energy Physics, 
ICHEP 06,\\
 26 July---2 August 2006, Moscow, Russia.

\end{center}

\vspace{1.0cm}
\begin{center}
{\em Stanford Linear Accelerator Center, Stanford University, 
Stanford, CA 94309} \\ \vspace{0.1cm}\hrule\vspace{0.1cm}
Work supported in part by Department of Energy contract DE-AC03-76SF00515.
\end{center}

\newpage
}

\begin{center}
\small

The \babar\ Collaboration,
\bigskip

%% author list as of 01-Jul-2006 (596 authors)
%
{B.~Aubert,}
{R.~Barate,}
{M.~Bona,}
{D.~Boutigny,}
{F.~Couderc,}
{Y.~Karyotakis,}
{J.~P.~Lees,}
{V.~Poireau,}
{V.~Tisserand,}
{A.~Zghiche}
\inst{Laboratoire de Physique des Particules, IN2P3/CNRS et Universit\'e de Savoie,
 F-74941 Annecy-Le-Vieux, France }
% ...

\end{center}\newpage

\section{Introduction}
\label{sec:Introduction}

 The precise measurement of \vub, the smallest element of the CKM 
matrix~\cite{CKM}, will strongly constrain the description of weak 
interactions and CP violation in the Standard Model.

The measurement of \vub\ requires the study of a $b \rightarrow u$ transition in 
a well-understood context. Semileptonic \ulnu\ decays (here, $\ell$ stands for 
an electron or a muon) are best for that purpose since they are much easier to 
understand theoretically than hadronic decays, and they are much easier to study 
experimentally than the less abundant purely leptonic decays.

For \pilnu\ decays,\footnote{Charge conjugation decays are implied throughout 
this paper.} the theoretical description of the quarks' strong interactions is 
parametrized by a single form factor, \fplus, where \qq\ is the squared invariant
mass of the $\ell^+ \nu$ system. Only the shape of \fplus\ can be measured 
experimentally. Its normalization is provided by theoretical calculations which 
currently suffer from relatively large uncertainties and, often, do not agree 
with each other. As a result, the normalization of the \fplus\ form factor is the
largest source of uncertainty in the extraction of \vub\ from the \pilnu\ 
branching fraction. Values of \fplus\ for \pilnu\ decays are provided by 
unquenched~\cite{HPQCD04, FNAL04} and quenched~\cite{APE} lattice QCD (LQCD)
calculations, presently reliable only at high \qq\ ($> 16$ \gevccsq), and by 
Light Cone Sum Rules calculations~\cite{ball04} (LCSR), based on approximations 
only valid at low \qq\ ($< 16$ \gevccsq), as well as by a quark 
model~\cite{isgw2}. The QCD theoretical predictions are at present more precise 
for \pilnu\ decays than for other exclusive \xulnu\ decays. Experimental data can
be used to discriminate between the various calculations by measuring the \fplus\
shape precisely, thereby leading to a smaller theoretical uncertainty on \vub.

  The present analysis of \pilnu\ decays aims to obtain an accurate measurement 
of the \fplus\ shape in order to extract a more precise value of \vub\ from the 
measurement of the total \pilnu\ branching fraction, \bfpilnu. To do so, we 
extract the \pilnu\ yields in \nQ\ bins of \qqr\ using a \lnrt\ and 
\qqr-dependent cuts. The quantity \qqr\ denotes the uncorrected measured value 
of \qq\ and will be referred to as ``raw''. The final \qq\ spectrum is 
corrected for reconstruction effects by applying an unfolding algorithm to the 
measured \qqr\ spectrum. The total \bfpilnu\ is given by the sum of the partial 
branching fractions \bfpilnuq. The \qq\ shape of the \fplus\ form factor is 
obtained from the normalized partial branching fractions \bfpilnuqp\ spectrum 
combined with two covariance matrices (one for the statistical errors and one for
the systematic errors) which give the correlations among the values of 
\bfpilnuqp\ measured in the different \qq\ bins. The measured \bfpilnuqp\ 
spectrum is fitted to a model-dependent parametrization~\cite{BK} of the \fplus\ 
form factor. The model-independent \bfpilnuqp\ spectrum and its correlation 
matrices are given explicitly to allow future studies with different \fplus\ 
parametrizations using the present data. The value of the CKM matrix element 
\vub\ is then derived from the form factor calculations, combined with the 
measured \DBR\s.

 The main innovation of this analysis is the use of a \lnrt\ which yields a 
much higher signal reconstruction efficiency than in past measurements, while
keeping the systematic errors at a relatively low level. The higher yield allows 
the utilization of a large number of \qq\ bins and the determination of the background 
composition using several independent fit parameters. In addition, we use 
\qqr-dependent cuts and estimate the \fplus\ shape systematic error.

 The data used in this analysis were collected with the \babar\ detector at the 
\pep2\ asymmetric $e^+e^-$ collider. The \babar\ detector is described 
elsewhere~\cite{ref:babar}. The following samples are used: 206.4 \invfb\ 
integrated luminosity of data collected at the \FourS\ resonance, corresponding 
to 227.4 million $B\bar{B}$ decays; 27.0 \invfb\ integrated luminosity of data 
collected approximately 40 \mev\ below the \FourS\ resonance (denoted 
``off-resonance data'' hereafter); standard \babar\ Monte Carlo (MC) simulation 
using GEANT4~\cite{geant} and EvtGen~\cite{evtgen}; 1.64 million \pilnu\ signal 
events using the FLATQ2 generator~\cite{bad809}; 2.02 billion generic \bbz\ and 
\bpbm\ events, and 2.44 billion generic \udsc\ and \tautau\ ``continuum'' events.

\section{Analysis Method}
\label{sec:Analysis}

Values of \vub\ have previously been extracted from \pilnu\ measurements by
CLEO~\cite{CLEOpilnu1, CLEOpilnu2}, \babar~\cite{PRDJochen, BAD1380} and 
BELLE~\cite{BELLEPiRholnu}. Our analysis is based on a novel technique denoted 
``\lnr''. The main motivation for implementing this technique is to maximize the 
extracted \pilnu\ signal yields in order to measure the \fplus\ shape as 
precisely as possible.

 Even though $B$ mesons are always produced in pairs at the \FourS\ resonance, 
a major feature of the neutrino reconstruction technique is that the decay of the
non-signal $B$ is not reconstructed. Instead, the signal $B$ mesons are directly 
identified using the measured $\pi^{\pm}$ and $\ell^{\mp}$ tracks together with 
the events' missing momentum as an approximation to the signal neutrino 
momentum~\cite{CLEOpilnu1, CLEOpilnu2, PRDJochen}. The neutrino four-momentum, 
$P_{miss}=(|\vec{p}_{miss}|,\vec{p}_{miss})$, is inferred from the difference 
between the four-momentum of the colliding-beam particles, 
$P_{\rm beams}=(E_{\rm beams},\vec{p}_{\rm beams})$, and the sum of the 
four-momenta of all charged and neutral particles detected in a single 
interaction, $P_{tot}=(E_{tot},\vec{p}_{tot})$, such that 
$\vec{p}_{miss}=\vec{p}_{\rm beams}-\vec{p}_{tot}$. Compared with the tagged 
analyses described in Refs.~\cite{BAD1380, BELLEPiRholnu}, the neutrino 
reconstruction approach yields a lower signal purity but a significant increase 
in the signal reconstruction efficiency. The new approach used in the \lnr\ 
further increases this efficiency compared with the previous untagged 
analyses~\cite{CLEOpilnu1, CLEOpilnu2, PRDJochen} by avoiding neutrino quality 
cuts (for example, a tight cut on the invariant missing mass to ensure the 
neutrino properties are well taken into account). Such cuts were required to 
allow the calculation of $\tilde{q}^2=(P_{\ell}+P_{\nu})^2$. To obtain the values
of \qqr, we use instead the neutrino-independent relation: 
$\tilde{q}^2=(P_{B}-P_{\pi})^2$. Although this relation is strictly true and 
Lorentz invariant, it cannot be used directly because the value of $P_B$ is not 
known. Only the value of $P_{\pi}$, measured in the laboratory frame, and that of
the \FourS\ 4-momentum are known. Nevertheless, since the $B$ momentum is small 
in the \FourS\ frame, a common approximation is to boost the pion to the \FourS\ 
frame and use the relation $\tilde{q}^2=(P_{B}-P_{\pi})^2$ in that frame, where 
the $B$ meson is assumed to be at rest. 

  However, a more accurate value of \qqr\ can be obtained in the so-called 
$Y$-average frame~\cite{YAve, DstrFF} where the pseudo-particle $Y$ has a 
4-momentum defined by $P_Y \equiv (P_{\pi}+P_{\ell})$. The angle $\theta_{BY}$ 
between the directions of the $p^*_B$ and $p^*_Y$ momenta\footnote{All variables 
denoted with an asterisk (e.g. $p^*$) are given in the \FourS\ rest frame; all 
others are given in the laboratory frame.} in the \FourS\ rest frame can be 
determined assuming energy-momentum conservation in a semileptonic $B 
\rightarrow Y \nu$ decay. Its value is given by: \cosBYDef, where $m_B, m_Y, 
E^*_B$, $E^*_Y, \vec p^{\,*}_B$ and $\vec p^{\,*}_Y$ refer to the masses, 
energies and momenta of the $B$ meson and the $Y$ ``particle''. Thus, in the 
\FourS\ frame, a cone is defined whose axis is given by the direction of the $Y$ 
momentum with the half-angle subtended at the apex given by $\theta_{BY}$. The 
apex corresponds to the vertex formed by the $Y$ and $B$ momenta directions. The 
$B$ momentum lies somewhere on the surface of the cone and thus its position is 
known only up to an azimuthal angle $\phi$ defined with respect to the $Y$ 
momentum. The value of \qqr\ in the $Y$-average frame is computed as follows: it 
first assumes that the $B$ rest frame is located at an arbitrary angle $\phi_0$, 
and the value of $q_0^2$ is calculated in that specific frame position. The 
values of $q^2_1$, $q^2_2$ and $q^2_3$ are then calculated with the $B$ rest 
frame at $\phi_1 = \phi_0 + 90^o$, $\phi_2 = \phi_0 + 180^o$ and $\phi_3 = 
\phi_0 + 270^o$, respectively. The value of \qqr\ in the $Y$-average frame is 
then defined as $\tilde{q}^2 = \frac{q^2_0 + q^2_1 + q^2_2 + q^2_3}{4}$. Using 
more than four values of $\phi_i$ does not significantly improve the \qqr\ 
resolution.

\begin{figure}[!htb]
\begin{center}
\epsfig{file=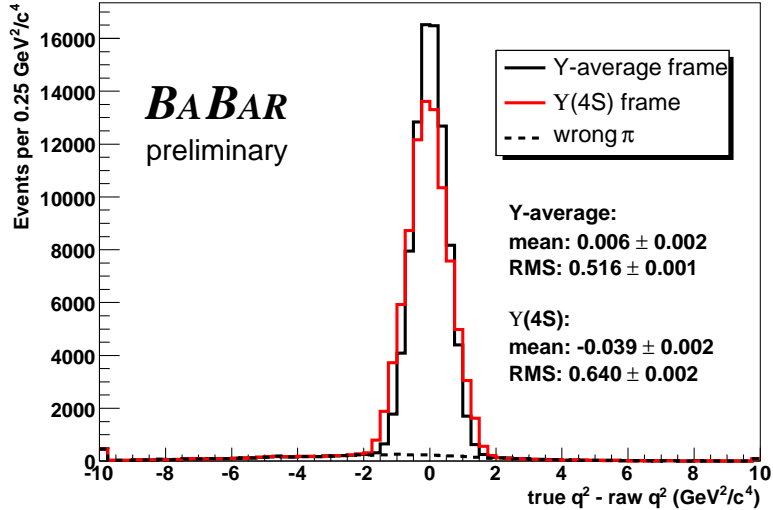,height=7cm}
\caption[$q^2$ resolution of \pilnu\ signal obtained in the Y-average and \FourS\
frames.]
{\label{figAveBetter} 
$q^2$ resolution of \pilnu\ signal events obtained in the Y-average and \FourS\ 
frames after all analysis cuts and MC corrections. The very long tail arises when
a track coming from the non-signal $B$ is wrongly selected as the signal pion. 
The numbers of entries in the first and last bins correspond to the sum of all 
entries with $\Delta q^2 < -9.75$ \gevccsq\ and $\Delta q^2 > 9.75$ \gevccsq, 
respectively.}
\end{center}
\end{figure}

 The use of the $Y$-average frame yields a \qq\ resolution that is approximately
20\% better than what is obtained in the usual \FourS\ frame where the $B$ meson 
is assumed to be at rest. We get an unbiased \qq\ resolution of 0.52 \gevccsq
when the selected pion candidate comes from a \pilnu\ decay (Fig. 
\ref{figAveBetter}), which accounts for approximately 91\% of our signal 
candidates after all the analysis selections. When a track from the non-signal 
$B$ is wrongly selected as the signal pion, the \qq\ resolution becomes very poor
and biased. 

\begin{figure}[!htb]
\begin{center}
\includegraphics[height=7cm]{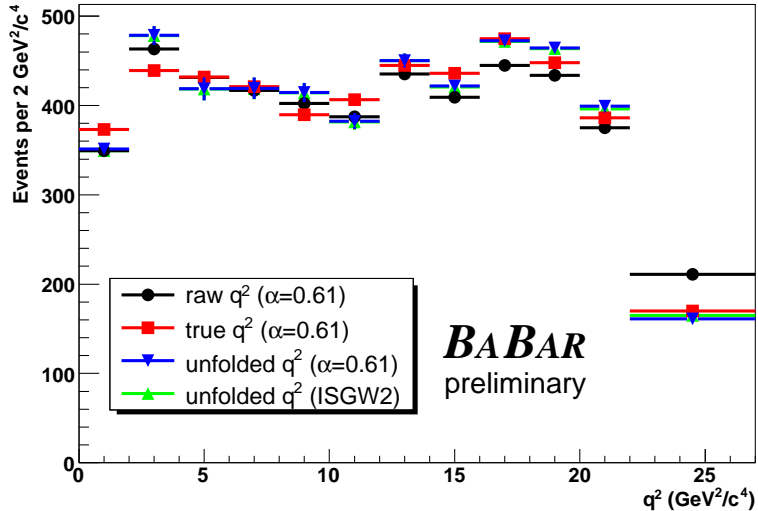}
\caption[Validation of the \qq-unfolding procedure with statistically independent
signal MC samples.]
{\label{q2UnfVsRec} 
Validation of the \qq-unfolding procedure. The true and raw yield distributions 
differ considerably for various values of \qq. However, independently of the 
signal generator used to compute the detector response matrix, the unfolded 
values of \qq\ match the true values within the statistical uncertainties of the 
unfolding procedure.}
\end{center}
\end{figure}

  We correct for our imperfect \qq\ resolution with a \qq-unfolding algorithm. 
This algorithm was validated with statistically independent signal MC samples. 
After all selections, the total signal MC sample contains approximately 120000
events. Five thousand such signal events were used to produce the raw \qqr\ and 
true \qq\ histograms. The remaining signal events were used to build the two 
unfolding matrices, using the simulated signal events reweighted~\cite{bad809} 
either to reproduce the \fplus\ shape measured in Ref.~\cite{PRDJochen} or with  
the weights calculated in Ref.~\cite{isgw2}. As illustrated in Fig. 
\ref{q2UnfVsRec}, the true and raw yield distributions differ considerably for 
various values of \qq. However, the unfolded values of \qq\ match the true values
within the statistical uncertainties of the unfolding procedure, independently of
the signal generator used to compute the detector response matrix. This shows 
that the \qq-unfolding procedure works as expected.
 
\begin{figure}[!htb]
\begin{center}
\includegraphics[height=10cm]{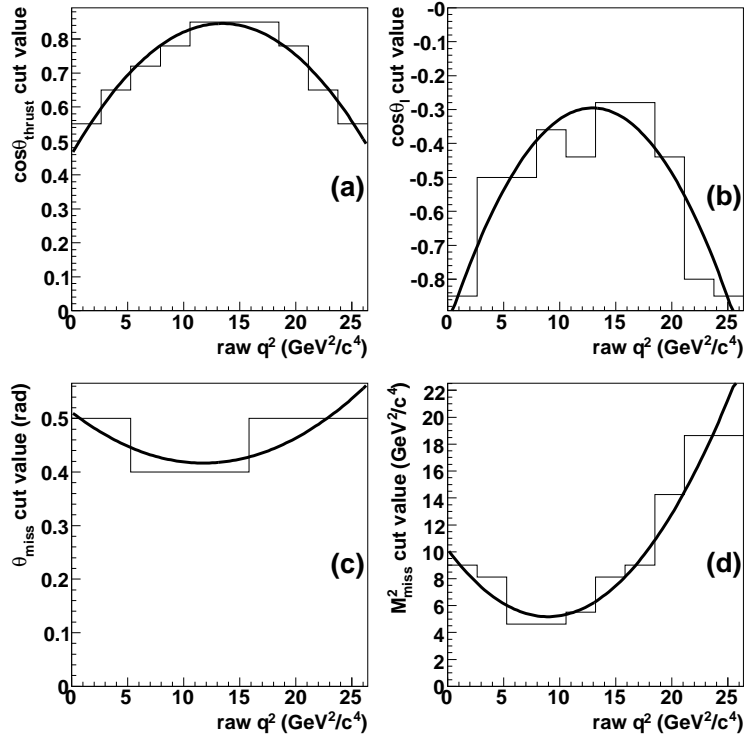}
\caption[Optimization of the $\cos\theta_{thrust}$, $\cos\theta_{\ell}$, 
$\theta_{miss}$ and $M^2_{miss}$ cuts as a function of \qqr.] {\label{qqCuts} 
Cut functions used in the analysis, for the variables $\cos\theta_{thrust}$ (a), 
$\cos\theta_{\ell}$ (b), $\theta_{miss}$ (c) and $M^2_{miss}$ (d).
The solid curves show the cut values as a function of \qqr. The histograms show 
the statistically optimal cut values obtained in 10 \qqr~bins based on the 
minimization of the quantity $\sqrt{(S+B)}/S$ in the \DeltaE-\mes\ signal region,
where $S$ represents the simulated signal yield and $B$ stands for the simulated 
background yield.}
\end{center}
\end{figure}

 To separate the \pilnu\ signal from the backgrounds, we require two well 
reconstructed tracks that fulfill tight lepton and pion identification criteria. 
The electron (muon) tracks are required to have a momentum greater than 0.5 (1.0)
\gevc\ in the laboratory frame. We do not cut on the pion momentum because it is 
very strongly correlated with \qq. The kinematic compatibility of the lepton and 
pion with a real \pilnu\ decay is constrained by requiring that a geometrical 
vertex fit~\cite{TreeFitter} of the two tracks gives a $\chi^2$ probability 
greater than 0.01, and by requiring that $-1 < \cos\theta_{BY} < 1$. Note that 
cuts whose values depend on the measured value of \qqr\ (Fig. 3) give the best
background rejection. Non-\BB events are suppressed by several conditions: we 
require at least four charged tracks in each event; we require the ratio of the 
second to the zeroth Fox-Wolfram moments~\cite{FoxWolfram} to be less than 0.5; 
we require the cosine of the angle between the $Y$'s thrust axis and the rest of 
the event's thrust axis, $\cos\theta_{thrust}$, to satisfy the 
relation\footnote{In the following relations, \qqr\ is given in units of 
\gevccsq.} $\cos\theta_{thrust} < 0.460 + 0.0576 \cdot \tilde{q}^2 - 0.00215 
\cdot \tilde{q}^4$ (Fig. \ref{qqCuts}); we require the polar angle associated 
with $\vec{p}_{miss}$ to satisfy the relation 2.7 rad $> \theta_{miss} > 
(0.512 - 0.0162 \cdot \tilde{q}^2 + 0.000687 \cdot \tilde{q}^4)$ rad 
(Fig. \ref{qqCuts}). Radiative Bhabha events are rejected using the criteria 
given in Ref.~\cite{BhabhaVeto} and photon conversion events are vetoed. Finally,
although the shapes of the \qqr, \DeltaE\ and \mes\ distributions in 
off-resonance data are very well reproduced by MC simulation in all lepton 
channels, there is an excess of nearly a factor of two in the yield values 
observed in data compared to the simulation in the electron/positron channels. We
then require 
$\frac{\vec{p}_{tot}\cdot\hat{z}}{E_{tot}} < 0.64$ and 
$\frac{\vec{p}_{tot}\cdot\hat{z}}{E_{tot}} > 0.35$ for candidates in the electron
and positron channels, respectively, where the $z$ axis is given by the electron 
beam direction~\cite{ref:babar}. This reduces the observed excess by removing
additional radiative Bhabha events as well as ``two-photon'' processes which are 
not included in the simulated continuum. 

\begin{figure}[!htb]
\begin{center}
\includegraphics[height=7cm]{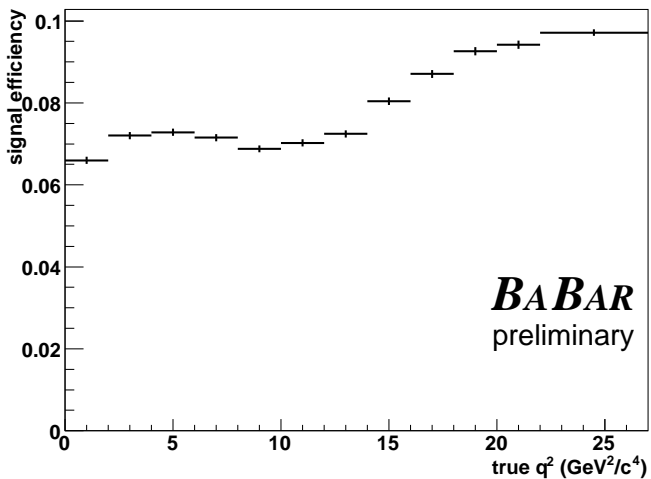}
\caption[Signal event reconstruction efficiency as a function of true $q^2$.]
{\label{figSigEff} 
Signal efficiency as a function of true $q^2$.}
\end{center}
\end{figure}

 To reject background $B\bar{B}$ events, we require the Y candidates to have 
$\cos\theta_{\ell} < 0.85$ and $\cos\theta_{\ell} > -0.938 + 0.0994 \cdot 
\tilde{q}^2 - 0.00384 \cdot \tilde{q}^4$ (Fig. \ref{qqCuts}), where 
$\theta_{\ell}$ is the helicity angle of the W boson~\cite{Singleton} 
reconstructed in the Y-average frame approximation. We reject $\jpsi \rightarrow
\mu^+ \mu^-$ decays, which can often be mistaken for \pimunu\ 
decays\footnote{This requirement is not necessary in the electron channel since 
the fake rate of charged pions by electrons is extremely low.}, by removing
candidates with $3.07 < m_Y < 3.13$ \gevcc. Of all the neutrino quality cuts 
utilized in Refs.~\cite{CLEOpilnu1, CLEOpilnu2, PRDJochen}, only the loose 
\qqr-dependent criterion on the squared invariant mass of $P_{miss}$ is used: 
$M^2_{miss}<(10.2 - 1.12 \cdot \tilde{q}^2 + 0.0625 \cdot \tilde{q}^4)$ \gevccsq\
(Fig. \ref{qqCuts}). We discriminate against the remaining backgrounds using the 
variables \DeltaEDef and 
$\ensuremath{m_{\rm ES} = \sqrt{(s/2+\vec{p}_B \cdot \vec{p}_{\rm beams})^2
/E_{\rm beams}^2- \vec{p}_B^{\,2}}} + (5.29 \gevcc\ - \sqrt{s}/2)$, where 
$\sqrt{s}$ is the total energy in the \FourS\ center-of-mass frame. Only 
candidates with \FitRegion\ are retained. When several candidates remain in an 
event after the above cuts, we select the candidate with $\cos\theta_{\ell}$ 
closest to zero and reject the others. This rejects 30\% of the combinatorial 
signal candidates while conserving 97\% of the correct ones and reduces the 
sensitivity of our analysis to the simulation of the candidates' multiplicity. 
After all cuts, the total signal event reconstruction efficiency varies between 
6.6\% and 9.7\%, depending on the \qq\ bin, as shown in Fig. \ref{figSigEff}. 

\begin{figure}[!htb]
\begin{center}
\includegraphics[height=12cm]{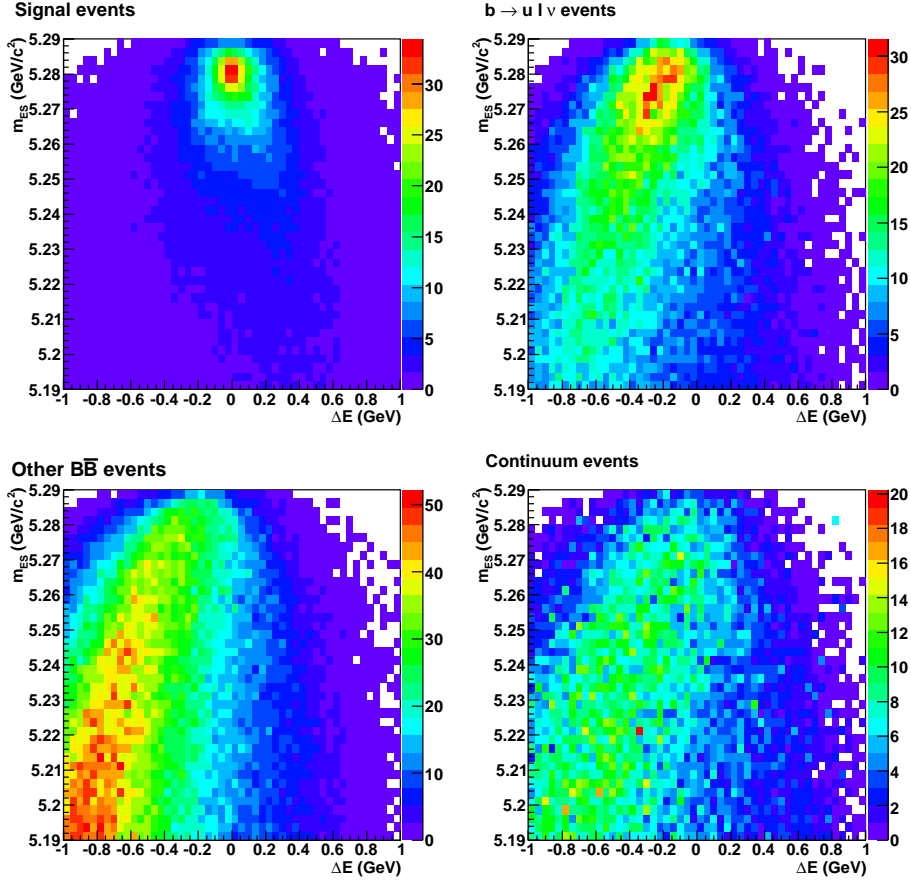}
\caption[\DeltaE-\mes\ distributions for the four types of events used in the 
signal extraction fit after all selections.]
{\label{dEmES} Sum of \DeltaE-\mes\ distributions over all \nQ\ bins of \qqr\ for
the four types of events used in the signal extraction fit after all cuts.}
\end{center}
\end{figure}

\begin{figure}[!htb]
\begin{center}
\includegraphics[height=7cm]{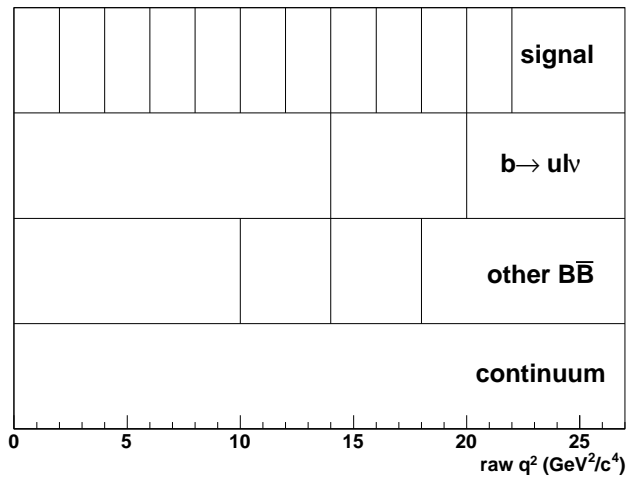}
\caption[\qqr\ binning used in the nominal fit.]
{\label{IllustrationQ2Bins} \qqr\ binning used in the fit of the MC PDFs to the
experimental data.}
\end{center}
\end{figure}

\begin{figure}
\begin{center}
\vspace{-2cm}
\epsfig{file=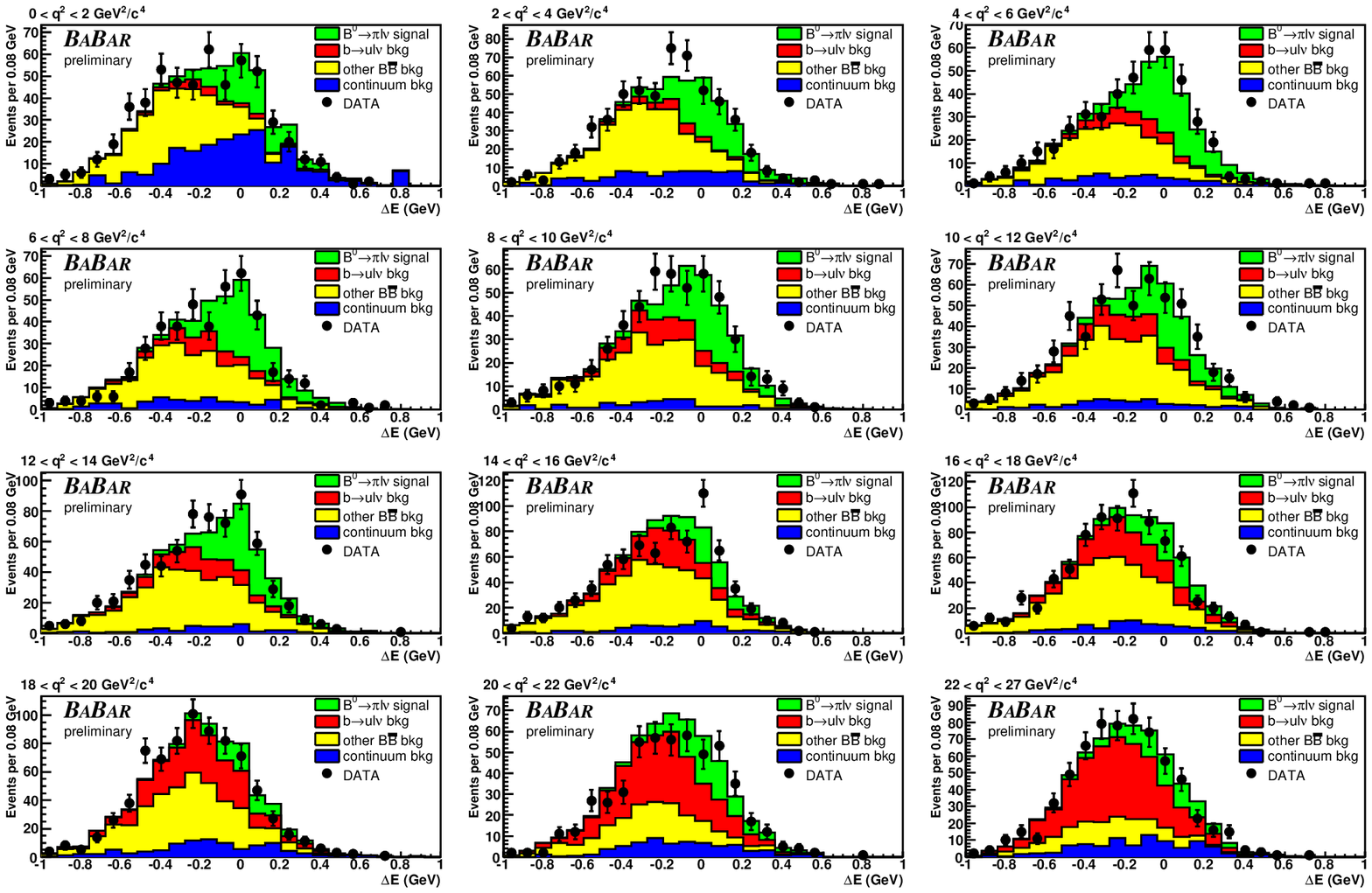,height=15cm,angle=90}
\caption[\DeltaE\ yield fit projections obtained in \nQ~\qqr~bins from the fit to
the experimental data, using the full \DeltaE-\mes\ fit region.]
{\label{dEFitProjMC} \DeltaE\ yield fit projections obtained in \nQ~\qqr~bins 
from the fit to the experimental data, using the full \DeltaE-\mes\ fit region. 
However, the displayed distributions are those after the cut: $m_{ES}>5.272$ 
\gevcc.}
\end{center}
\end{figure}

\begin{figure}
\begin{center}
\vspace{-2cm}
\epsfig{file=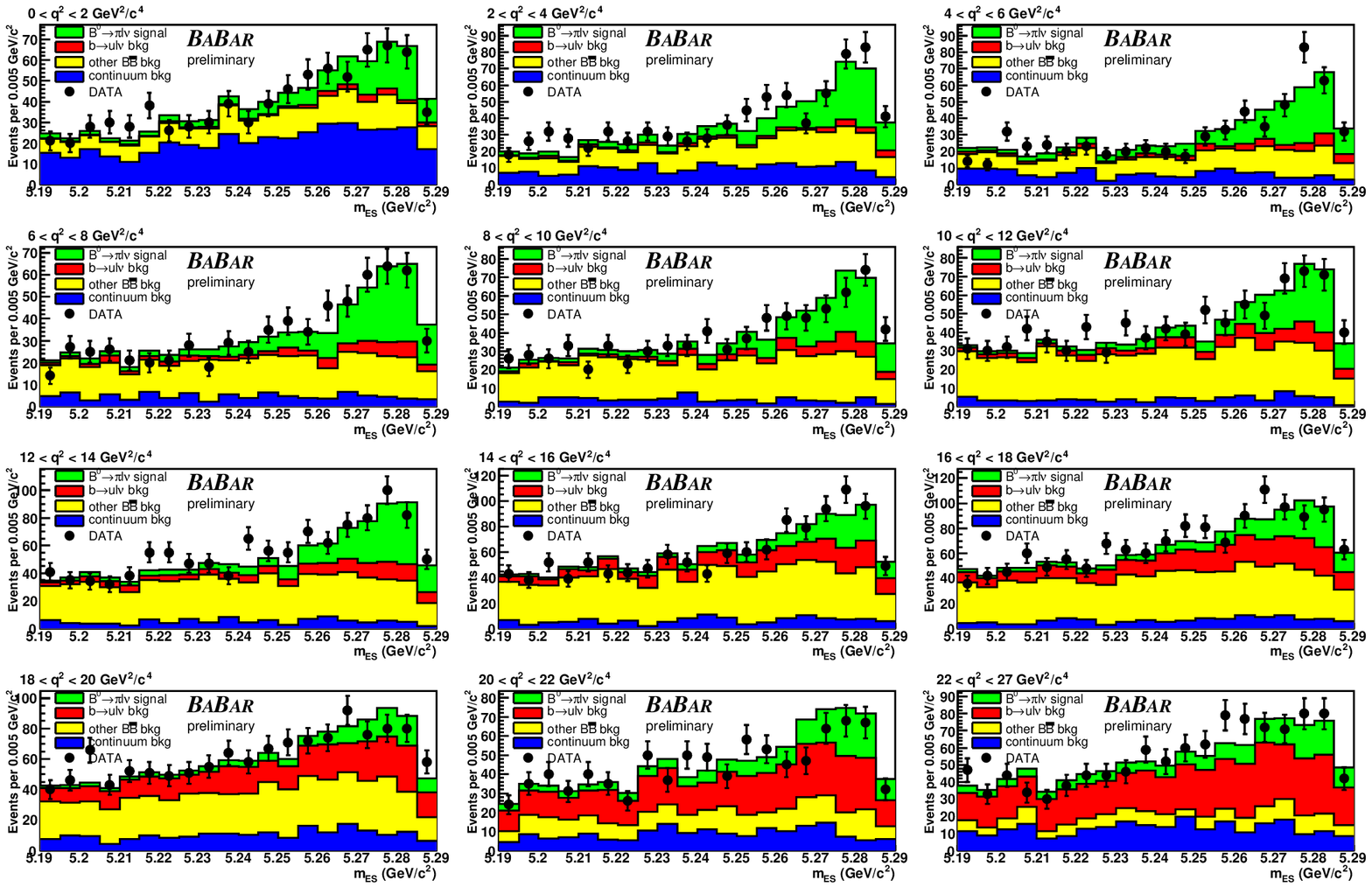,height=15cm,angle=90}
\caption[\mes\ yield fit projections obtained in \nQ~\qqr~bins from the fit to 
the experimental data, using the full \DeltaE-\mes\ fit region.]
{\label{mESFitProjMC} \mes\ yield fit projections obtained in \nQ~\qqr~bins from 
the fit to the experimental data, using the full \DeltaE-\mes\ fit region. 
However, the displayed distributions are those after the cut: 
$\vert\Delta E\vert<0.18$ \gev.}
\end{center}
\end{figure}

 To obtain the \pilnu\ signal yield in each of the \nQ\ reconstructed \qqr\ 
bins, we perform a 2+1 dimensional (\DeltaE-\mes, \qq) extended binned maximum 
likelihood fit based on a method developed by Barlow and Beeston~\cite{Barlow}. 
The fitted data samples in each \qqr\ bin are divided into four categories: 
\pilnu\ signal and three backgrounds, \ulnu, \obb, and continuum. The distinct 
structure of these four types of events in the 2-dimensional \DeltaE-\mes\ plane 
is illustrated in Fig. \ref{dEmES}. Since the correlation between \DeltaE\ and 
\mes\ cannot be neglected and is difficult to parametrize, we use the 
\DeltaE-\mes\ histograms obtained from the MC simulation as two-dimensional 
probability density functions (PDF). The simulated signal events are 
reweighted~\cite{bad809} to reproduce the \fplus\ shape measured in 
Ref.~\cite{PRDJochen}. The \qqr\ shape of the simulated non-\BB\ continuum 
background is scaled to match the off-resonance data control sample containing 
both $e^{\pm}$ and $\mu^{\pm}$ events, while the scaling of the yields requires 
separate $e^{\pm}$ and $\mu^{\pm}$ samples. The fit of the MC PDFs to the 
experimental data gives the values of twenty parameters: twelve parameters for 
the twelve signal \qqr\ bins, three for the \ulnu\ background, four for the \obb\
background, and one for the continuum background, as illustrated in Fig. 
\ref{IllustrationQ2Bins}. The number and type of fit parameters were chosen to 
provide a good balance between reliance on simulation predictions, complexity of 
the fit and total error size. The corresponding \DeltaE\ and \mes\ fit 
projections in each \qqr\ bin for the experimental data are shown in Figs. 
\ref{dEFitProjMC} and \ref{mESFitProjMC}. We obtain a total signal yield of 
\sigY\ events, while for backgrounds the \ulnu\ yield is \ulnY\ events, the 
\obb\ yield is \obbY\ events, and the continuum yield is \conY\ events. The fit 
has a $\chi^2$ value of 428/388 degrees of freedom. In the more restricted signal
region ($\mes > 5.272$ \gevcc, $|\DeltaE| < 0.18$ \gev), the total signal yield 
is $1340 \pm 40$ events and the total background yield $2527 \pm 55$ events, for 
a signal/background ratio of $0.53 \pm 0.02$.  

 From the raw signal yields, the unfolded partial branching fractions \bfpilnuq\ 
are calculated using the inverted detector response matrix given by the 
simulation and the signal efficiencies. The total branching fraction \bfpilnu\ is
given by the sum of the partial \bfpilnuq\ thereby greatly reducing the 
sensitivity of the total branching fraction to the uncertainties of the \fplus\ 
form factor values, which have a small but non-negligeable effect on the values 
of the efficiencies. 

 To reduce the uncertainties in evaluating the \fplus\ shape, instead of fitting
the measured \bfpilnuq\ spectrum, we fit the normalized distribution, \bfpilnuqp,
obtained by dividing the measured $\DBR$ spectrum by the measured value of the 
total branching fraction. With this approach, a number of correction factors 
cancel out, leading to a significant decrease in the systematic error. We fit the
\bfpilnuqp\ spectrum using a PDF based on the $f^+(q^2,\alpha)$ parametrization 
of Becirevic-Kaidalov~\cite{BK}, which is proportional to the standard 
differential decay rate for a semileptonic $B$ decay to a pseudo-scalar meson 
($X_{PS}$): 

\begin{equation}
\label{eqDiffDecRate}
F(q^2,\alpha) = \frac{|\vec{p_{\pi}}|^3\cdot |f^{+}(q^2,\alpha)|^2}
{\int_0^{q_{max}}|\vec{p_{\pi}}|^3\cdot |f^{+}(q^2,\alpha)|^2~dq^2}
~~\propto \frac{d\Gamma(B\rightarrow X_{PS}~\ell^{+}\nu_{\ell})}{dq^2}
\end{equation} 
where $|\vec{p_{\pi}}|= \sqrt{\frac{(m_B^2+m_{\pi}^2-q^2)^2}{4m_B^2}-m_{\pi}^2}$,
and the $f^{+}(q^2,\alpha)$ function is:
\begin{equation}
f^{+}(q^2,\alpha) = \frac{f_0}{(1-q^2/m^2_{B^*})\cdot(1-\alpha q^2/m^2_{B^*})}
\end{equation}

\noindent
The value of $f_0$ cancels out in Eq. 1. Note that the data can also be used to 
extract the \fplus\ shape parameter(s) using \textit{any} theoretical 
parametrization, e.g. those of Refs.~\cite{HPQCD04, Hill, BZ}. The $\chi^2$ value
minimized in the fit is defined in terms of the covariance matrix $U$ to take 
into account the correlations between the measurements in the various \qq\ bins:

\begin{equation}
\label{eqChi2}
\chi^2 =\sum_{q_i,q_j} \left(\DBR(\ensuremath{q^2_i})/\BR -
\int_{q^2_i} F(q^2)dq^2\right)
~U^{-1}_{ij}~\left(\DBR(\ensuremath{q^2_j})/\BR\ -\int_{q^2_j} F(q^2)dq^2\right),
\end{equation}

\noindent
where $\int_{q^2_i} F(q^2)dq^2$ denotes the integral of Eq. \ref{eqDiffDecRate} 
over the range of the i$^{th}$ \qq\ bin and $\sum_i \int_{q^2_i} F(q^2)dq^2 
\equiv 1$. The central value of the parameter $\alpha$, and its total error, are 
obtained using the total covariance matrix in Eq. 3. In the present case, in 
which the errors on \bfpilnuqp\ are more or less uniform across the $q^2$ bins, 
using the statistical or the systematic covariance matrix in Eq. 3 yields the 
statistical or the systematic errors for $\alpha$, respectively. Their quadratic 
sum is in fact consistent with the total error. The statistical covariance matrix
is given directly by the fit to the signal. The systematic and total covariance 
matrices are obtained as described in the next section. 

\section{Systematic Error Studies}
\label{sec:Systematics}

Numerous sources of systematic uncertainties have been considered. Their values
are established by a procedure in which variables used in the analysis are
varied within their allowed range, generally established in previous \babar\ 
analyses. For the uncertainties due to the detector simulation, the variables are
the tracking efficiency of all charged tracks (varied between $\pm$ 0.7\% and 
$\pm$ 1.4\%), the particle identification efficiencies of signal candidate tracks
(varied between $\pm$ 0.2\% and $\pm$ 2.2\%), the calorimeter efficiency (used 
in the full-event reconstruction, and varied between $\pm$ 0.7\% and $\pm$ 1.8\% 
for photons, and up to $\pm$ 25\% for \KL mesons) and the energy deposited in the
calorimeter by \KL mesons (varied up to $\pm$ 15\%). For the uncertainties due to
the generator-level inputs to the simulation, the variables are the branching 
fractions of the background processes \ulnu, \clnu\ and $D \rightarrow K^0_L X$ 
as well as the branching fraction of the $\Upsilon(4S) \rightarrow B^0\bar{B^0}$
decay (all varied within their known errors~\cite{PDG04} except when the 
branching fractions have not been measured. In those cases, the branching 
fractions are varied by $\pm$ 100\% from their presumed central values). The 
\rholnu\ form factors are varied within bounds of $\pm$ 10\% at \qq\ $=0$ and 
$\pm$ 16\% at $q^2_{max}$, given by recent Light-Cone Sum Rules 
calculations~\cite{ball05}, while the \dstrlnu\ form factors are varied within 
their measured uncertainties~\cite{DstrFF}, between $\pm$ 5.5\% and $\pm$ 9.6\%. 
To take into account an additional subtle effect on the uncertainty of the signal
efficiency, the \pilnu\ form factor shape parameter $\alpha$ is varied between 
its recently measured central value~\cite{PRDJochen} and that of the unquenched 
HPQCD calculations~\cite{HPQCD04}, a difference of 0.2. Finally, for the 
uncertainties due to the modelling of the continuum, there are variations in the 
continuum yields and in the \qqr, $\Delta E$ and $\ensuremath{m_{\rm ES}}$ 
shapes, as discussed in Section 2.

The systematic errors are then given by the variation in the final values of the 
branching fractions when the data are re-analyzed with different values of the 
simulation parameters. For each source of uncertainty, we generate at least one 
hundred MC samples in which the simulation parameters are varied according to a 
Gaussian standard deviation. This standard deviation is given by the range of the
variations listed above. For each MC sample, the entire analysis is reproduced 
leading to new signal efficiencies, \qq-unfolding matrices, \DeltaE-\mes\ PDFs 
and \pilnu\ signal yields from a fit to the same data sample. The rms value 
of the resulting branching fraction distribution is taken to be the value of the 
systematic error contributed by the source of uncertainty under study. The 
individual branching fractions are also used to generate two-dimensional 
\bfpilnuqi\ versus \bfpilnuqj\ distributions, for all $(q^2_i,q^2_j)$ 
combinations. The linear correlation coefficient in each of these distributions 
is used to build the covariance matrix of the \bfpilnuqq\ measurements for each 
source of systematic error. The total systematic covariance matrix is then simply
given by the sum of all the individual covariance matrices and is used to 
calculate the total systematic error on the branching fraction. The same 
procedure is repeated for the normalized branching fractions. The resulting total
systematic covariance matrix yields in this case the total systematic error on 
the parameter $\alpha$. All the statistical and systematic uncertainties are 
given in Tables A-1 and A-2 of Appendix A while the correlation matrices of the 
normalized branching fractions are presented in Tables A-3 and A-4. \\

\section{Results}
\label{sec:Physics}

The values of the partial \bfpilnuq\ and total \bfpilnu\ branching fractions are 
given in Table A-1, those of the normalized partial \bfpilnuqp\ branching 
fractions are listed in Table A-2. In Table A-1, we also give the small 
uncertainties on the signal efficiency and \qq-unfolding matrix due to the signal
MC statistics. The total branching fraction error is due in large part to the 
photon and tracking efficiency systematic uncertainties. However, the use of the
\lnr\ did indeed reduce their impact~\cite{PRDJochen}. The systematic errors 
arising from the branching fractions and form factors of the backgrounds have 
been greatly reduced by the many-parameter fit to the background yields in the 
twelve bins of \qqr. As expected, the errors on \bfpilnuqp\ are mostly 
statistical. The value of the total \bfpilnu\ branching fraction obtained from 
the sum of the partial \bfpilnuq\ branching fractions is:
\begin{center}
  \bfpilnu\ $= (1.44~\pm~0.08_{stat}~\pm~0.10_{syst}) \times 10^{-4}$ 
\end{center}

\begin{figure}[!htb]
\begin{center}
\epsfig{file=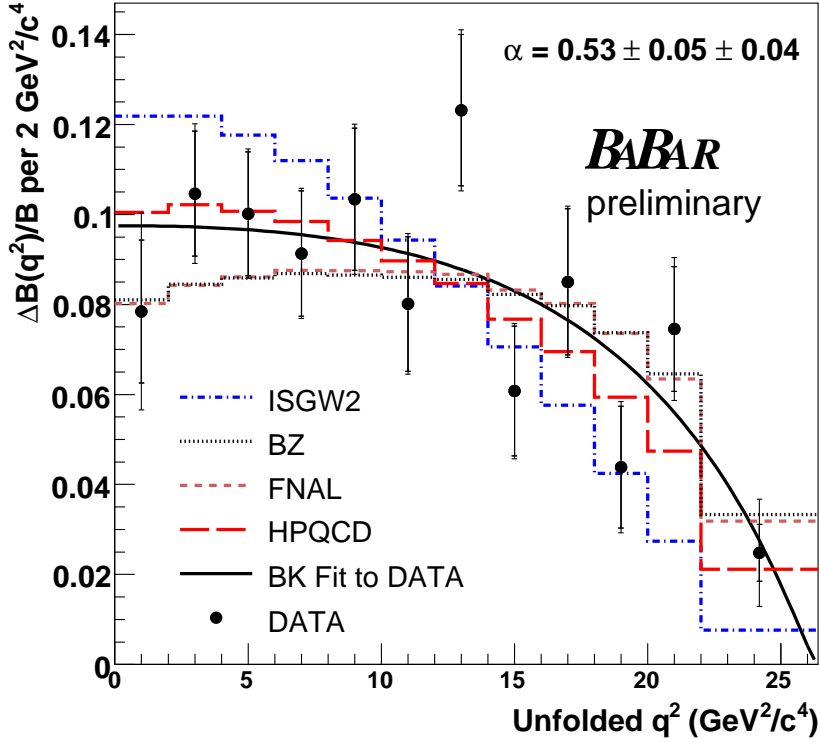,height=11cm}
\caption[Differential decay rate formula fitted to the normalized partial 
\bfpilnuqp\ spectrum obtained in the MC simulation.]
{\label{figFplus} 
Differential decay rate formula (Eq. \ref{eqDiffDecRate}) fitted to the 
normalized partial \bfpilnuqp\ spectrum in 12 bins of \qq. The smaller error 
bars are statistical only while the larger error bars include statistical 
and systematic uncertainties. The BK parametrization (solid black curve) 
reproduces the data quite well ($\chi^2=8.8$ for 11 degrees of freedom) with the
parameter $\alpha=0.53 \pm 0.05 \pm 0.04$. The data are also compared to LCSR 
calculations~\cite{ball04} (dotted line), unquenched LQCD 
calculations~\cite{HPQCD04} (long dashed line),~\cite{FNAL04} (short dashed line)
and the ISGW2 quark model~\cite{isgw2} (dash-dot line).}  
\end{center}
\end{figure}

\begin{table}[h]
\caption[$\chi^2$ values and associated probabilities for various QCD calculation
predictions.]{
\label{chi2Table} $\chi^2$ values and associated probabilities for various QCD 
calculation predictions and for the ISGW2 model compared to our measured \qq\ 
spectrum, for twelve degrees of freedom.}
\begin{center}
\begin{tabular}{|c|c|c|c|c|}
\hline\hline
      & \multicolumn{2}{c|}{stat error only} &
\multicolumn{2}{c|}{stat+syst errors} \\
QCD calculation & $\chi^2$ & $Prob(\chi^2)$ (\%) & $\chi^2$ & $Prob(\chi^2)$ (\%)
\\ \hline
ISGW2~\cite{isgw2} & 49.5 & $<$ 0.01 & 34.1 & 0.07 \\
Ball-Zwicky~\cite{ball04} & 17.0 & 14.9 & 13.0 & 37.2 \\
FNAL~\cite{FNAL04} & 16.2 & 18.2 & 12.5 & 41.0 \\
HPQCD~\cite{HPQCD04} & 14.2 & 28.6 & 10.2 & 60.2 \\
\hline\hline
\end{tabular}
\end{center}
\end{table}

The normalized \bfpilnuqp\ distribution is displayed in Fig.~\ref{figFplus}
together with the result of a \fplus\ shape fit using the BK parametrization and
theoretical predictions. We obtain a value of $\alpha=0.53 \pm 0.05 \pm 0.04$. In
Table \ref{chi2Table}, we give the $\chi^2$ values and their associated 
probabilities for the four different calculations. These values were obtained by 
comparing, bin by bin, the data with the central values of the form-factor 
calculations, ignoring the theoretical errors. Our experimental data are clearly
incompatible with the ISGW2 quark model. A more definitive choice among the 
remaining theoretical calculations must await a substantial increase in 
statistics. 

 Monte Carlo studies have shown that there is no significant fit bias for the 
\bfpilnu\ but a small 3.8\% bias in the \fplus\ parameter $\alpha$ which has been
incorporated in its systematic error. Various cross-checks have also been 
performed. The results were obtained separately for the electron and muon decay 
channels, for the off-resonance data replacing the continuum PDF, for the 
different \DeltaE-\mes\ and \qqr\ binnings, and for the variations of all the
analysis cuts, one at a time. All the cross-check studies were found to be 
consistent with the final results. 

  We extract $|V_{ub}|$ from the partial branching fractions \DBR\ using
$|V_{ub}| = \sqrt{\DBR/(\tau_B^0\Delta \zeta)}$, where $\tau_B^0 = (1.536 
\pm 0.014)$ ps~\cite{PDG04} is the $B^0$ lifetime and $\Delta \zeta = 
\Delta \Gamma/|V_{ub}|^2$ is the normalized partial decay rate predicted by 
various form factor calculations. We use the LCSR calculations for $q^2 < 16$ 
\gevccsq\ and the LQCD calculations for $q^2 > 16$ \gevccsq. The results are
shown in Table \ref{vubtable}. The uncertainties of the form-factor normalization
are taken from Refs.~\cite{HPQCD04, FNAL04, APE, ball04}. We obtain values of 
$|V_{ub}|$ ranging from $3.6\times 10^{-3}$ to $4.1 \times 10^{-3}$. For the most
recently published unquenched LQCD calculation~\cite{HPQCD04}, we obtain 
$|V_{ub}| = \left(4.1 \pm 0.2_{stat} \pm 0.2_{syst}\pmff{0.6}{0.4} \right)
\times {10^{-3}}$.  

\begin{table}[h]
\caption[Values of $|V_{ub}|$ derived from the form-factor calculations. The 
first two errors arise from the statistical and systematic uncertainties of the 
partial branching fractions. The third error comes from the uncertainties on
$\Delta \zeta$ due to the models.]{\label{vubtable}Values of $|V_{ub}|$ derived 
from the form-factor calculations. The first two errors arise from the 
statistical and systematic uncertainties of the partial branching fractions. The 
third error comes from the uncertainties on $\Delta \zeta$ due to the 
theoretical calculations.} 
\begin{center}
\begin{tabular}{|c|c|c|c|c|}
\hline\hline
 & $q^2$ (\gevccsq) &$\DBR$ ($10^{-4}$) & $\Delta\zeta$ ($ps^{-1}$) & $|V_{ub}|$ 
($10^{-3}$)\\ \hline
Ball-Zwicky~\cite{ball04} & $< 16$ & $1.07~\pm~0.06~\pm~0.08$ & $5.44~\pm~1.43$ &
 $3.6~\pm~0.1~\pm~0.1~{}^{+0.6}_{-0.4}$\\
HPQCD~\cite{HPQCD04}      & $> 16$ & $0.37~\pm~0.04~\pm~0.03$ & $1.46~\pm~0.35$ &
 $4.1~\pm~0.2~\pm~0.2~{}^{+0.6}_{-0.4}$\\
FNAL~\cite{FNAL04}        & $> 16$ & $0.37~\pm~0.04~\pm~0.03$& $1.83~\pm~0.50$ & 
$3.6~\pm~0.2~\pm~0.2~{}^{+0.6}_{-0.4}$\\
APE~\cite{APE}            & $> 16$ & $0.37~\pm~0.04~\pm~0.03$& $1.80~\pm~0.86$ & 
$3.7~\pm~0.2~\pm~0.2~{}^{+1.4}_{-0.7}$\\
\hline\hline
\end{tabular}
\end{center}
\end{table}

\section{Summary}
\label{sec:Summary}

 The succesful development of the \lnrt\ shows that it is not always necessary to
have very pure signal samples to control the systematic errors. The gain in 
statistical precision can overcome the negative features of large backgrounds. 
This technique could thus be used advantageously in future measurements, possibly
those of other exclusive \xulnu\ decays.

 In the present analysis, we have obtained the total \pilnu\ branching fraction 
from the values of the partial branching fractions measured in 12 bins of \qq\
and the \fplus\ shape parameter using the Becirevic-Kaidalov parametrization. We 
summarize these results in Table \ref{summary} together with the value of 
$|V_{ub}|$ extracted from a recent calculation of the form factor~\cite{HPQCD04}.

\begin{table}[h]
\begin{center}
\caption[Summary of the main results.]
{\label{summary}Summary of the main results.} 
\begin{tabular}{l}
\\ \hline\hline
\bfpilnu\ $= (1.44\pm0.08_{stat}\pm0.10_{syst}) \times 10^{-4}$ \\
$\alpha_{BK}=0.53\pm0.05_{stat}\pm0.04_{syst}$ \\
$|V_{ub}|=\left(4.1\pm0.2_{stat}\pm0.2_{syst}\pmff{0.6}{0.4}\right) 
\times 10^{-3}$ \\
\hline\hline
\end{tabular}
\end{center}
\end{table}

  Our value for the \bfpilnu\ is the most precise measurement to date and, by
itself, is of comparable precision to the current world average~\cite{HFAG}:
\bfpilnu\ $= (1.34\pm0.08_{stat}\pm0.08_{syst}) \times 10^{-4}$. The
new value of the BK parameter is an improvement over our previous measurement 
$\alpha = 0.61 \pm 0.09$~\cite{PRDJochen} (no systematic error quoted). 

  The errors in Table A-2 together with the correlation matrices of the 
statistical and systematic errors presented in Tables \ref{StatCovMC} and 
\ref{SystCovMC} will allow the present data to be studied with different future 
\fplus\ parametrizations. A simple $\chi^2$ calculation already shows that our 
data are incompatible with the predictions of the ISGW2 quark model. 

\section{Acknowledgements}
\label{sec:Acknowledgments}

We are grateful for the 
extraordinary contributions of our \pep2\ colleagues in
achieving the excellent luminosity and machine conditions
that have made this work possible.
The success of this project also relies critically on the 
expertise and dedication of the computing organizations that 
support \babar.
The collaborating institutions wish to thank 
SLAC for its support and the kind hospitality extended to them. 
This work is supported by the
US Department of Energy
and National Science Foundation, the
Natural Sciences and Engineering Research Council (Canada),
Institute of High Energy Physics (China), the
Commissariat \`a l'Energie Atomique and
Institut National de Physique Nucl\'eaire et de Physique des Particules
(France), the
Bundesministerium f\"ur Bildung und Forschung and
Deutsche Forschungsgemeinschaft
(Germany), the
Istituto Nazionale di Fisica Nucleare (Italy),
the Foundation for Fundamental Research on Matter (The Netherlands),
the Research Council of Norway, the
Ministry of Science and Technology of the Russian Federation, and the
Particle Physics and Astronomy Research Council (United Kingdom). 
Individuals have received support from 
the Marie-Curie IEF program (European Union) and
the A. P. Sloan Foundation.

\vspace{5mm}
\noindent
\large
{\bf{Appendix A}}
\normalsize
\vspace{5mm}

 The values of the partial \bfpilnuq\ and total \bfpilnu\ branching fractions are
given in Table A-1, those of the normalized partial \bfpilnuqp\ branching 
fractions are listed in Table A-2. All the statistical and systematic 
uncertainties as well as their associated correlation matrices are given in 
Tables A-1, A-2, A-3 and A-4. 

\renewcommand{\thetable}{A-\arabic{table}}
\setcounter{table}{0}

\begin{sidewaystable}[f]
\caption[Partial \bfpilnuq\ and total \bfpilnu\ $(\times 10^7)$ and their errors 
$(\times 10^7)$ from all sources.]
{\label{BFErrTable} Partial \bfpilnuq\ and total \bfpilnu\ $(\times 10^7)$ and 
their errors $(\times 10^7)$ from all sources.}
\begin{small}
\begin{center}
\begin{tabular}{|c||c|c|c|c|c|c|c|c|c|c|c|c||c|c|c|}
\hline
\qq\ intervals (\gevccsq) & 0-2 & 2-4 & 4-6 & 6-8 & 8-10 & 10-12 & 12-14 & 14-16 & 16-18 & 18-20 & 20-22 & 22-26.4 & Total & \qq$<$16 & \qq$>$16 \\ \hline
fitted BF             & 113.2& 151.1& 144.6& 131.8& 149.3& 115.7& 177.8& 87.7& 122.7& 63.3& 107.6& 78.7& 1443.6& 1071.2& 372.4  \\ \hline\hline
fitted yield stat err & 22.9& 20.0& 19.9& 20.1& 22.8& 21.5& 24.3& 20.8& 23.4& 19.5& 20.0& 20.1& 83.4& 63.3& 44.1 \\ \hline
trk eff               & 14.7& 1.5& 6.1& 3.3& 3.7& 3.4& 4.4& 4.0& 2.4& 4.6& 1.3& 4.4& 40.3& 39.7& 4.5 \\ 
$\gamma$ eff          & 15.2& 1.0& 5.3& 7.0& 3.3& 9.0& 7.3& 4.3& 3.2& 4.0& 2.8& 1.9& 56.9& 50.9& 6.7 \\ 
\KL\ eff \& E         & 1.6& 1.0& 1.1& 1.4& 1.5& 0.8& 2.2& 1.6& 1.7& 1.5& 1.2& 1.8& 7.1& 5.2& 3.7 \\ 
Y PID \& trk eff      & 2.3& 2.9& 2.1& 2.8& 2.3& 1.7& 3.4& 1.5& 2.3& 0.9& 1.9& 1.8& 22.1& 15.8& 6.5 \\ 
\hline
continuum yield       & 3.2& 0.6& 0.4& 0.1& 0.1& 0.2& 0.2& 0.6& 0.6& 1.3& 0.8& 2.1& 4.2& 2.2& 4.6 \\ 
continuum \qqr        & 12.9& 2.3& 1.8& 1.2& 1.5& 1.3& 1.1& 2.0& 2.2& 3.6& 3.9& 8.4& 8.8& 8.8& 12.4 \\ 
continuum \mes        & 6.1& 0.6& 0.4& 0.1& 0.9& 0.6& 0.1& 0.7& 0.5& 1.1& 1.2& 2.0& 12.4& 7.7& 4.7 \\ 
continuum \DeltaE     & 2.6& 2.2& 0.6& 1.4& 2.4& 0.5& 0.4& 1.0& 0.7& 1.4& 3.4& 3.3& 17.8& 9.2& 8.6 \\ 
\hline
$D\rightarrow K^0_L$ BF& 8.1& 6.3& 8.2& 3.3& 3.9& 4.5& 3.4& 4.6& 5.5& 4.9& 4.3& 4.5& 51.6& 34.2& 17.5 \\ 
\clnu\ BF             & 3.0& 2.6& 1.5& 2.8& 6.2& 1.2& 4.3& 2.0& 1.7& 2.7& 1.5& 1.6& 17.0& 13.2& 5.0 \\ 
\ulnu\ BF             & 1.7& 1.8& 1.1& 1.0& 1.6& 1.6& 2.0& 1.5& 1.7& 2.9& 7.9& 7.5& 16.6& 9.6& 11.5 \\ 
$\Upsilon(4S)\rightarrow B^0\bar{B^0}$ BF& 2.3& 3.3& 2.3& 2.0& 2.1& 1.9& 2.9& 1.1& 2.0& 0.8& 2.4& 1.4& 23.9& 17.7& 6.3 \\ 
\hline
\dstrlnu\ FF          & 1.7& 1.3& 0.2& 2.0& 4.0& 1.0& 2.6& 0.9& 1.0& 0.8& 0.8& 2.5& 12.5& 10.2& 2.7 \\ 
\rholnu\ FF & 4.0& 1.2& 3.4& 1.7& 1.1& 1.7& 2.6& 3.9& 1.3& 1.9& 1.6& 3.5& 18.3& 14.4& 5.7\\
\pilnu\ FF            & -1.2& -0.0& 0.3& 0.1& 0.3& 0.0& 0.7& -0.3& -0.2& -1.3& 1.6& 4.5& 4.7& 0.0& 4.7 \\ 
\hline
signal MC stat error  & 1.8& 2.6& 2.4& 2.6& 2.4& 2.3& 2.6& 1.5& 1.7& 1.1& 1.3& 1.1& 5.5& 5.1& 2.2 \\ 
B counting            & 1.2& 1.7& 1.6& 1.5& 1.6& 1.3& 2.0& 1.0& 1.4& 0.7& 1.2& 0.9& 15.9& 11.8& 4.1 \\ 
\hline\hline
total syst error          & 28.0& 9.8& 13.1& 10.6& 11.3& 11.7& 12.5& 9.7& 8.8& 10.3& 11.8& 15.6& 102.7& 83.0& 31.3 \\ \hline\hline
total error & 36.2& 22.3& 23.8& 22.7& 25.5& 24.5& 27.3& 23.0& 25.1& 22.1& 23.3& 25.4& 132.3& 104.4& 54.1\\ \hline
\end{tabular}
\end{center}
\end{small}
\end{sidewaystable}

\begin{sidewaystable}[f]
\caption[Normalized partial \bfpilnuqp\ ($\times 10^3$) and their errors ($\times
10^3$).]{\label{ShapeErrTable} 
Normalized partial \bfpilnuqp\ ($\times 10^3$) and their errors ($\times 10^3$).}
\begin{small}
\begin{center}
\begin{tabular}{|c||c|c|c|c|c|c|c|c|c|c|c|c|}
\hline
\qq\ intervals (\gevccsq) & 0-2 & 2-4 & 4-6 & 6-8 & 8-10 & 10-12 & 12-14 & 1
4-16  & 16-18 & 18-20 & 20-22 & 22-26.4 \\ \hline
normalized partial BF & 78.4& 104.6& 100.2& 91.3& 103.4& 80.1& 123.2& 60.8& 85.0&
 43.9& 74.5& 54.6 \\ \hline\hline
fitted yield stat err & 15.9& 13.9& 13.8& 13.9& 15.8& 14.9& 16.8& 14.4& 16.2& 
13.5& 13.9& 14.0 \\ \hline
signal MC stat error  & 1.3& 1.8& 1.6& 1.8& 1.7& 1.6& 1.8& 1.0& 1.2& 0.8& 0.9& 
0.8 \\ 
track eff             & 7.7& 2.3& 1.6& 0.8& 1.0& 0.6& 1.0& 1.2& 2.2& 2.1& 2.3& 
4.3 \\ 
photon eff          & 7.0& 3.4& 0.7& 2.1& 1.9& 3.0& 1.0& 0.8& 1.6& 1.2& 4.1& 1.8 
\\ 
\KL\ eff \& E         & 0.9& 0.9& 0.6& 0.9& 0.9& 0.4& 1.4& 1.1& 1.0& 1.0& 0.6& 
1.2 \\ 
Y PID \& trk eff      & 2.3& 0.5& 0.2& 0.7& 0.3& 0.1& 0.7& 0.2& 0.4& 0.5& 0.2& 
0.5 \\ 
\hline
continuum yield       & 2.3& 0.3& 0.2& 0.2& 0.3& 0.1& 0.4& 0.3& 0.2& 0.8& 0.4& 
1.4 \\ 
continuum \qqr        & 8.9& 2.1& 1.6& 1.1& 0.6& 1.3& 0.7& 1.4& 1.1& 2.3& 2.3& 
5.8 \\ 
continuum \mes        & 3.4& 0.4& 0.5& 0.7& 1.4& 0.2& 1.0& 0.2& 0.4& 0.4& 0.2& 
0.9 \\ 
continuum \DeltaE\    & 0.9& 0.3& 0.8& 0.2& 0.5& 1.3& 1.8& 0.2& 0.7& 0.5& 1.5& 
1.6 \\ 
\hline
\BR($D\rightarrow K^0_L$) & 2.9& 3.3& 2.3& 1.9& 2.2& 2.6& 3.2& 2.5& 1.5& 2.3& 
1.3& 1.3 \\ 
\BR(\clnu)             & 1.5& 2.8& 1.4& 1.3& 3.3& 1.0& 2.1& 1.3& 1.1& 1.6& 1.0& 
0.8 \\ 
\BR(\ulnu)             & 0.6& 1.3& 0.6& 0.9& 1.2& 0.7& 1.7& 0.6& 1.6& 2.0& 4.8& 
4.9 \\ 
\BR($\Upsilon(4S)\rightarrow B^0\bar{B^0}$) & 0.3& 0.6& 0.2& 0.2& 0.4& 0.1& 0.3& 
0.3& 0.1& 0.6& 0.4& 0.2 \\ 
\hline
\dstrlnu\ FF          & 0.6& 1.7& 0.8& 0.7& 2.0& 0.5& 0.8& 0.6& 1.3& 0.7& 0.3& 
1.3 \\ 
\rholnu\ FF & 1.8& 0.6& 1.4& 0.6& 0.7& 0.3& 3.2& 2.0& 0.7& 1.4& 1.5& 1.8\\
\pilnu\ FF            & 1.1& 0.3& 0.1& 0.2& 0.1& 0.2& 0.1& 0.4& 0.4& 1.0& 0.9& 2.9 \\ 
\hline
\hline
total syst error          & 15.1& 7.1& 4.5& 4.3& 5.7& 4.9& 6.4& 4.4& 4.5& 5.5& 
7.8& 10.1 \\ \hline\hline
total error & 21.9& 15.6& 14.5& 14.6& 16.8& 15.7& 18.0& 15.1& 16.9& 14.6& 15.9& 
17.2 \\ \hline
\end{tabular}
\end{center}
\end{small}
\end{sidewaystable}

\begin{table}[f]
\caption[Correlation matrix of the normalized partial \bfpilnuqp\
statistical
errors.]{\label{StatCovMC}Correlation matrix of
the normalized partial \bfpilnuqp\ statistical errors.}
\begin{center}
\begin{small}
\begin{tabular}{r|cccccccccccc}
\qq\ intervals \\
(\gevccsq) & 0-2 & 2-4 & 4-6 & 6-8 & 8-10 & 10-12 & 12-14 & 14-16 & 16-18
& 18-20 & 20-22 & 22-26.4\\ \hline
0-2& 1.00 & -0.20 & 0.12 & -0.00 & -0.01 & 0.04 & 0.04 & -0.01 & -0.00 &
0.01 & 0.01 & 0.01\\
2-4& -0.20 & 1.00 & -0.32 & 0.14 & 0.03 & 0.01 & 0.02 & -0.01 & -0.00 &
-0.00 & 0.00 & 0.00\\
4-6& 0.12 & -0.32 & 1.00 & -0.31 & 0.20 & 0.05 & 0.13 & -0.02 & -0.00 &
-0.00 & 0.00 & 0.00\\
6-8& -0.00 & 0.14 & -0.31 & 1.00 & -0.22 & 0.14 & 0.08 & -0.01 & -0.00 &
-0.00 & -0.00 & -0.00\\
8-10& -0.01 & 0.03 & 0.20 & -0.22 & 1.00 & -0.23 & 0.19 & -0.03 & 0.00 &
-0.00 & -0.00 & -0.01\\
10-12& 0.04 & 0.01 & 0.05 & 0.14 & -0.23 & 1.00 & -0.02 & 0.02 & -0.00 &
0.00 & -0.00 & 0.00\\
12-14& 0.04 & 0.02 & 0.13 & 0.08 & 0.19 & -0.02 & 1.00 & -0.24 & 0.01 &
-0.02 & -0.00 & -0.00\\
14-16&-0.01 & -0.01 & -0.02 & -0.01 & -0.03 & 0.02 & -0.24 & 1.00 & 0.00 &
0.11 & -0.03 & -0.00\\
16-18& -0.00 & -0.00 & -0.00 & -0.00 & 0.00 & -0.00 & 0.01 & 0.00 & 1.00 &
0.01 & -0.04 & -0.01\\
18-20& 0.01 & -0.00 & -0.00 & -0.00 & -0.00 & 0.00 & -0.02 & 0.11 & 0.01 &
1.00 & -0.18 & -0.11\\
20-22&0.01 & 0.00 & 0.00 & -0.00 & -0.00 & -0.00 & -0.00 & -0.03 & -0.04 &
-0.18 & 1.00 & -0.01\\
22-26.4& 0.01 & 0.00 & 0.00 & -0.00 & -0.01 & 0.00 & -0.00 & -0.00 & -0.01
& -0.11 & -0.01 & 1.00\\
\end{tabular}
\end{small}
\end{center}
\end{table}

\begin{table}[f]
\caption[Correlation matrix of the normalized partial \bfpilnuqp\
systematic
errors.]{\label{SystCovMC}Correlation matrix of
the normalized partial \bfpilnuqp\ systematic errors.}
\begin{center}
\begin{small}
\begin{tabular}{r|cccccccccccc}
\qq\ intervals \\
(\gevccsq) & 0-2 & 2-4 & 4-6 & 6-8 & 8-10 & 10-12 & 12-14 & 14-16 & 16-18
& 18-20 & 20-22 & 22-26.4\\ \hline
0-2& 1.00 & -0.58 & 0.04 & -0.20 & -0.27 & 0.19 & -0.21 & 0.12 & -0.48 &
0.15 & -0.40 & -0.48\\
2-4& -0.58 & 1.00 & 0.09 & -0.16 & -0.01 & -0.35 & 0.12 & -0.16 & 0.38 &
-0.39 & 0.34 & 0.16\\
4-6& 0.04 & 0.09 & 1.00 & -0.02 & -0.26 & 0.16 & -0.25 & 0.12 & 0.06 &
-0.05 & -0.28 & -0.15\\
6-8& -0.20 & -0.16 & -0.02 & 1.00 & 0.28 & 0.46 & 0.33 & -0.05 & -0.03 &
-0.20 & -0.37& -0.11\\
8-10& -0.27 & -0.01 & -0.26 & 0.28 & 1.00 & -0.38 & 0.52 & -0.35 & 0.01 &
-0.33 & 0.15 & -0.06\\
10-12& 0.19 & -0.35 & 0.16 & 0.46 & -0.38 & 1.00 & 0.03 & 0.33 & -0.22 &
0.14 & -0.53 & -0.28\\
12-14& -0.21 & 0.12 & -0.25 & 0.33 & 0.52 & 0.03 & 1.00 & -0.53 & 0.12 &
-0.34 & 0.03 & -0.35\\
14-16& 0.12 & -0.16 & 0.12 & -0.05 & -0.35 & 0.33 & -0.53 & 1.00 & -0.33 &
0.36 & -0.17 & -0.10\\
16-18& -0.48 & 0.38 & 0.06 & -0.03 & 0.01 & -0.22 & 0.12 & -0.33 & 1.00 &
0.07 & 0.12 & 0.04\\
18-20& 0.15 & -0.39 & -0.05 & -0.20 & -0.33 & 0.14 & -0.34 & 0.36 & 0.07 &
1.00 & -0.29 & -0.04\\
20-22& -0.40 & 0.34 & -0.28 & -0.37 & 0.15 & -0.53 & 0.03 & -0.17 & 0.12 &
-0.29 & 1.00 & 0.17\\
22-26.4& -0.48 & 0.16 & -0.15 & -0.11 & -0.06 & -0.28 & -0.35 & -0.10 &
0.04 & -0.04 & 0.17 & 1.00\\
\end{tabular}
\end{small}
\end{center}
\end{table}

\end{document}